\documentclass[aps,prd,onecolumn,groupedaddress,showpacs,nofootinbib,amssymb
]{revtex4}
\usepackage[dvips]{graphicx}
\usepackage{amssymb}
\usepackage{amsmath}
\usepackage{graphicx}
\usepackage{amsfonts}
\usepackage{bm}
\begin{document}

\title{Some Classes of Gravitational Shock Waves from Higher Order Theories of Gravity}
\author{
V.K. Oikonomou,$^{1,2}$\,\thanks{v.k.oikonomou1979@gmail.com}}
\affiliation{$^{1)}$ Tomsk State Pedagogical University, 634061 Tomsk, Russia\\
$^{2)}$ Laboratory for Theoretical Cosmology, Tomsk State University of Control Systems
and Radioelectronics (TUSUR), 634050 Tomsk, Russia\\
}

\begin{abstract}
We study the gravitational shock wave generated by a massless high energy particle in the context of higher order gravities of the form $F(R,R_{\mu \nu}R^{\mu \nu},R_{\mu \nu \alpha \beta}R^{\mu \nu \alpha \beta})$. In the case of $F(R)$ gravity, we investigate the gravitational shock wave solutions corresponding to various cosmologically viable gravities, and as we demonstrate the solutions are rescaled versions of the Einstein-Hilbert gravity solution. Interestingly enough, other higher order gravities result to the general relativistic solution, except for some specific gravities of the form $F(R_{\mu \nu}R^{\mu \nu})$ and $F(R,R_{\mu \nu}R^{\mu \nu})$, which we study in detail. In addition, when realistic Gauss-Bonnet gravities of the form $R+F(\mathcal{G})$ are considered, the gravitational shock wave solutions are identical to the general relativistic solution. Finally, the singularity structure of the gravitational shock waves solutions is studied, and it is shown that the effect of higher order gravities makes the singularities milder in comparison to the general relativistic solutions, and in some particular cases the singularities seem to be absent.
 
\end{abstract}

%PACS numbers: 04.50.Kd, 95.36.+x, 98.80.-k, 98.80.Cq

\maketitle

%\makeatletter
%\renewcommand{\theequation}{\Roman{section}\,\Alph{subsection}.\arabic{equation}}
%\@addtoreset{equation}{subsection}{section}
%\makeatother

%\makeatletter
%\renewcommand{\theequation}{\Roman{section}.\arabic{equation}}
%\@addtoreset{equation}{section}
%\makeatother

\def\pp{{\, \mid \hskip -1.5mm =}}
\def\cL{\mathcal{L}}
\def\be{\begin{equation}}
\def\ee{\end{equation}}
\def\bea{\begin{eqnarray}}
\def\eea{\end{eqnarray}}
\def\tr{\mathrm{tr}\, }
\def\nn{\nonumber \\}
\def\e{\mathrm{e}}

\section{Introduction}

One of the most difficult challenges in theoretical physics is to find the ultimate quantum gravity theory describing the Universe at high energies. Up to date, many consistent aspects of a quantum cosmology theory have appeared in the literature, with the most promising being the Loop Quantum Cosmology theory \cite{LQC1,LQC3,LQC4,LQC5,LQC6,LQC7,LQC9,LQC10,LQC11,LQC12,LQC13,LQC14,LQC15}. However, even Loop Quantum Cosmology cannot be considered as a complete quantum gravity theory describing the Universe at high energies, since many theoretical questions need to be addressed.

The complete quantum gravity theory will describe gravitational and particle phenomena at high energies, and the high energy regime needs to modify in some way the background spacetime that the interaction takes place. An interesting gravitational solution that describes a massless particle moving with high energy was found some time ago by Aichelburg and Sexl and later developed by Dray and t'Hooft \cite{gsw1,gsw2}. This high energy gravitational solution was called a gravitational shock wave, and actually this solution describes the spacetime around a particle whose energy is dominated by kinetic energy rather that rest mast, and hence it is effectively massless. The gravitational shock wave can be considered as a promising way to reveal quantum gravitational phenomena, since this solution distorts the background spacetime of a massless particle with energy near or higher than the Planck scale. Various aspects of gravitational shock waves have been studied in the literature after the seminal work of Aichelburg and Sexl \cite{gsw1} and Dray and t'Hooft \cite{gsw2}, and for an incomplete list see for example \cite{gsw3,gsw4,gsw5,gsw6,gsw7,gsw8,gsw9,gsw10,gsw11,gsw12,gsw13,gsw14,gsw15,gsw16,gsw17,gsw18}. To our opinion, the feature that a highly energetic particle actually deforms the background spacetime is very appealing, and can potentially have applications in heavy ion collisions, see for example Ref. \cite{gsw9,gsw10}, or can also have applications at cosmological scales, see for example Ref. \cite{gsw11}.

With the gravitational shock wave being a very simple but potentially interesting doorway towards the quantum aspects of gravity, in this paper we shall calculate the gravitational shock wave solutions of various viable higher order gravities \cite{highord1,highord2,highord3,highord4,highord5}, focusing on cosmologically viable $F(R)$ gravities, but also other realistic gravities (see \cite{reviews1,reviews1a,reviews2,reviews3,reviews4} for reviews on higher order gravities and modified gravity). We shall be interested mainly in theories which the gravitational action contains functions of the form $F(R,\Psi,\Omega)$, with $R$ being the Ricci scalar, while $\Psi$ is the Ricci tensor squared scalar, that is $\Psi=R_{\mu \nu}R^{\mu \nu}$, and $\Omega$ is the Kretschmann scalar, that is, $\Omega=R_{\mu \nu \alpha \beta}R^{\mu \nu \alpha \beta}$. We will study the gravitational shock waves solutions for some classes of these higher order gravities, and as we will see, most of the solutions are similar to the general relativistic solution and only in two classes of models non-trivial solutions occur which are different from the general relativistic case. Some similar works on a specific class of higher order gravity was performed sometime ago, see  for example \cite{gsw14,gsw15}. Also we shall discuss another higher order gravity class of models, containing the Gauss-Bonnet scalar $\mathcal{G}$ \cite{gaussb1,gaussb2,gaussb3,gaussb4}.

The motivation for studying gravitational shock waves solutions in the context of higher order gravities comes mainly from the recent observational confirmation of gravitational waves from the LIGO collaboration \cite{LIGO}. Particularly, it is known that the linearization of higher order gravities leads to extra polarization modes corresponding to gravitational waves, among which are a spin 0 and spin 2 massive modes \cite{highord1}, which are ghost modes. The possibility of detecting some of these polarization modes of a stochastic gravitational wave renders the study of higher order gravities very timely in all contexts, since this would be a direct indication that the standard Einstein-Hilbert gravity should be enlarged \cite{ref11,ref12}. However the gravitational shock wave is not a usual gravitational wave, since the shock wave accompanies a highly energetic moving particle and thus it is not generated by extreme gravitational processes, but resides more to the quantum aspects of gravity phenomena. Apart from the above reasoning, the reason for studying higher order gravities comes from the fact that many paradigms coming from cosmology and quantum field theory suggest that the standard Einstein-Hilbert gravity has to be extended, for example the early and late-time era of the expanding Universe and the physics of extreme gravitational phenomena. With our study we aim to bridge two different kinds of theories, which may give some insights to the quantum aspects of gravity, and particularly we will study the gravitational shock waves generated by the classical higher order gravities. Thus in some sense, even we do not actually make use of any quantum gravity assumption, we study a quantum gravitational phenomenon in the context of classical gravitational theories, with quantum being justified by the assumption of a high energy particle.

The first class of higher order theories of gravity we shall study is the $F(R)$ gravity, and we focus on various viable models of $F(R)$ gravity which satisfy both astrophysical local constraints as well as global constraints \cite{reviews1,reviews1a,reviews2,reviews3,reviews4}. Particularly, we focus on several well-known viable models, and specifically we study an exponential model of $F(R)$ gravity which unifies the late and early-time acceleration era firstly studied in  \cite{frexponential}, and a variant form studied in \cite{oikofr}. In addition, we discuss the Hu-Sawicki model \cite{frhu}, the Appleby-Battye model \cite{frbattye}, and the Starobinsky model \cite{frstarob}. As we demonstrate, all these viable models have very similar gravitational shock wave solutions which are rescaled forms of the Einstein-Hilbert solution.

 After discussing the $F(R)$ cases, we proceed to discuss various higher order gravities of the form $F(R,\Psi,\Omega)$. As we show, most of the solutions are either similar to the general relativistic case and only two cases yield non-trivial results. The similarities of the resulting solutions is probably due to the specific form of the gravitational shock wave solution, and we discuss this issue in some detail. Finally, in the end we discuss in brief the case of a specific Gauss-Bonnet gravity. A general comment is that due to the form of the resulting Einstein equations, the form of the $F(R,\Psi,\Omega)$ is very much restricted, since it has to obey $F(0,0,0)=0$, as we show. Thus in some cases, unless a cosmological constant is included, the modified gravity is trivial.

This paper is organized as follows: In section II we present the essential features of the gravitational shock wave solution corresponding to the Einstein-Hilbert gravity. In section III we study many types of higher order gravity, starting from $F(R)$ gravity. We discuss the quantitative features of the gravitational shock wave solution corresponding to various viable $F(R)$ cosmological models. In addition, we study the higher order gravities of the form $F(R,\Psi)$ and we present the differences and similarities among the solutions we found. In the end of section III, we study in brief various realistic Gauss-Bonnet gravities of the form $R+F(\mathcal{G})$, and we compare the various gravitational shock waves we found. The conclusions follow in the end of the paper.

\section{Essential Features of Gravitational Shock Waves from Einstein-Hilbert Gravity}

In this section we describe the gravitational shock wave solution in the context of standard Einstein-Hilbert gravity. The gravitational shock wave solution was firstly discovered by Aichelburg and Sexl in \cite{gsw1}, where it was shown that the gravitational field of a massless highly energetic particle propagating in Minkowski space is a gravitational impulsive wave, which is also an asymmetric plane fronted gravitational wave. The spacetime metric is equal to,
\begin{equation}\label{specificmetric}
\mathrm{d}^2s=-\mathrm{d}u\mathrm{d}v+H(u,x,y)\mathrm{d}u^2+\mathrm{d}x^2+\mathrm{d}y^2\, ,
\end{equation}
with the coordinates $u$ and $v$ being equal to $u=t-z$ and $v=t+z$. For similar interesting metrics see \cite{hervik}. Also we shall assume that the function $H(u,x,v)$ is equal to $H(u,x,y)=G(x,y)\, \delta (u)$, so practically the gravitational shock wave is located at $u=0$ and the particle that generates this solution has momentum $p$ and moves along the $v$ direction, with the gravitational shock wave accompanying the particle. The geometric quantities corresponding to the metric (\ref{specificmetric}) are given in detail in the Appendix. The gravitational shock waves have the characteristic effect of a discontinuity $\Delta v$ at $u=0$ and also a refraction effect takes place, see for example \cite{gsw2} for details. The profile function $H(u,x,y)$ depends crucially on the source of the wave, which we will assume to be a massless particle with momentum $p$, and the only non-zero component of the energy momentum tensor is $T_{uu}=p\,\delta(x,y) \delta (u)$. The Einstein-Hilbert equations are of course $G_{\mu \nu}=8\pi G T_{\mu \nu}$, so by using the non-vanishing components of the Ricci tensor, and the fact that the Ricci scalar $R$ is zero for the metric (\ref{specificmetric}), the resulting Einstein equations are,
\begin{equation}\label{einteincaseequation}
\left (\frac{\partial^2}{\partial x^2}+\frac{\partial^2}{\partial y^2}\right )G_{EH}(x,y)=-16\pi G p\,\delta(x,y)\, ,
\end{equation}
with the solution being the well known result \cite{gsw1,gsw2},
\begin{equation}\label{einsteinsolution}
G_{EH}(x,y)=-4\,G\,p\,\ln \left( \frac{r^2}{r_0^2}\right)\, ,
\end{equation}
with $r_0$ being an integration constant having the same dimensions as $r$, and $r$ is the radial coordinate in the $(x,y)$ plane, that is $r=\sqrt{x^2+y^2}$. Note we used the notation $G_{EH}(x,y)$ for the Einstein-Hilbert profile function solution to discriminate from the other cases we study later. Notice that the profile function and therefore the gravitational wave is singular at the origin $(x,y)=(0,0)$, an issue which we shall discuss in the next sections, and we compare the resulting higher order profile functions to the Einstein-Hilbert profile (\ref{einsteinsolution}). By looking the metric of the gravitational shock wave (\ref{specificmetric}) it can be seen that it is relatively simple, which an intriguing feature to see how this metric becomes in the case that the theory is a higher order effective theory of gravity. In this way we will grasp the quantum effects of a highly energetic moving massless particle in flat space, even though using a classical effective gravitational theory. The qualitative features of such a theory can provide us with useful information about these quantum aspects of classical gravitational extensions of Einstein-Hilbert gravity. In the next section we shall investigate how the profile function $G(x,y)$ becomes in the case of higher order gravity theories, by using the same assumption of a massless point-like spin-less source propagating in Minkowski spacetime.

\section{Gravitational Shock Waves from Higher Order Gravities}

In this section we study the gravitational shock wave solutions for various higher order gravity theories with gravitational action,
\begin{equation}\label{actionhigherorder}
\mathcal{S}=\int \mathrm{d}^4x\sqrt{-g}F(R,\Psi,\Omega)+\mathcal{S}_m\, ,
\end{equation}
where $R$ is the Ricci scalar, and $\Psi$ and $\Omega$ are the Ricci tensor squared scalar, that is $\Psi=R_{\mu \nu}R^{\mu \nu}$, and the Kretschmann scalar, that is, $\Omega=R_{\mu \nu \alpha \beta}R^{\mu \nu \alpha \beta}$, respectively, and also $\mathcal{S}_m$ is the matter action containing the sources of the gravitational field, which corresponds to an energy momentum tensor $T_{\mu \nu}$. In the context of the metric formalism, by varying the action (\ref{actionhigherorder}) with respect to the metric tensor $g_{\mu \nu}$, we obtain the following equations of motion \cite{highord1,highord2,highord3,highord4,highord5},
\begin{align}\label{generalequationsofmotion}
& F_RG_{\mu \nu}=\frac{1}{2}g_{\mu \nu}(F-R\,F_R)-(g_{\mu \nu}\square -\nabla_{\mu}\nabla_{\nu})F_R-2(F_{\Psi}R^{\alpha}_{\mu}R_{\alpha \nu}+F_{\Omega}R_{\alpha b c \mu}R^{abc}_{\nu})\\ \notag &-g_{\mu \nu}\nabla_a\nabla_b(F_{\Psi} R^{ab})-\square (F_{\Psi}R_{\mu \nu})+
2\nabla_a\nabla_b (F_{\Psi}R^{\alpha}\, _{(\mu}\delta^b_{\nu )}+2F_{\Omega}R^{\alpha}_{(\mu \nu )}\, ^{b})+8\pi G T_{\mu \nu}\, ,
\end{align} 
where $F_R=\frac{\partial F}{\partial R}$, $F_{\Psi}=\frac{\partial F}{\partial \Psi}$, $F_{\Omega}=\frac{\partial F}{\partial \Omega}$, and $\square = g^{ab}\nabla_a \nabla_b$ is the d'Alembert operator. In the following sections we shall consider various higher order gravities. As we will see, there are striking similarities between the gravitational shock waves solutions of various higher order gravities. These similarities are due to the fact that the Ricci scalar $R$, the Ricci squared scalar $\Psi$ and the Kretschmann scalar $\Omega$, are equal to zero for the metric (\ref{specificmetric}) as it can be seen in the Appendix.

\subsection{Solutions from Cosmologically Viable $F(R)$ Gravities}

In the literature there exist various cosmologically viable $F(R)$ gravity models which we mentioned in the introduction, which satisfy very stringent constraints in order for these to be considered as viable, see for example the reviews \cite{reviews1,reviews1a,reviews2,reviews3,reviews4}. These models satisfy the local astrophysical constraints and also the global constraints, and in addition these have quite appealing features, since they describe successfully the early and late-time acceleration eras, and also yield quite interesting astrophysical solutions. The reader is referred to the informative reviews \cite{reviews1,reviews1a,reviews2,reviews3,reviews4} for details.

In our analysis we shall seek for gravitational shock waves solutions for several well known viable models. Particularly, we consider the exponential model of Ref. \cite{frexponential}, in which case the $F(R)$ gravity is,
\begin{equation}\label{expon}
F(R)=R-2\Lambda (c-e^{-R/R_0}),\,\,\, R_0,\Lambda ,c>0\, ,
\end{equation}
and a variant form of this exponential model, studied in Ref. \cite{oikofr}, in which case, the $F(R)$ gravity function is,
\begin{equation}\label{oikoexp}
F(R)=R-\frac{1}{A+Be^{-R/D}}+\frac{C}{A+B}\, ,
\end{equation}
and as we will see shortly, these two models have quite similar gravitational shock wave solutions. To this end, we focus on the model (\ref{expon}) and we present only the result of the model (\ref{oikoexp}). In the $F(R)$ gravity case, the equations of motion become quite simpler and these read,
\begin{align}\label{frofmotion1}
& F_RG_{\mu \nu}=\frac{1}{2}g_{\mu \nu}(F-R\,F_R)-(g_{\mu \nu}\square -\nabla_{\mu}\nabla_{\nu})F_R\, ,
\end{align}
and by using the fact that the Ricci scalar is zero for the metric (\ref{specificmetric}), these result to the following equation,
\begin{equation}\label{expodiff1}
(1-\frac{2\Lambda}{R_0})R_{\mu \nu}-\frac{1}{2}g_{\mu \nu}2 (1-c)\Lambda =8\pi GT_{\mu \nu}\, .
\end{equation}
For the metric (\ref{specificmetric}) the only non-zero component is $R_{uu}=-\frac{1}{2}\left (\frac{\partial^2}{\partial x^2}+\frac{\partial^2}{\partial y^2}\right )H(u,x,y)$, and also the source of the gravitational field is a highly energetic particle with momentum $p$, located at $u=0$, so the energy momentum tensor has only one non-zero component $T_{uu}=p\,\delta(x,y) \delta (u)$. Hence, the only non-trivial components of the Einstein equations are the $(u,u)$ and $(x,x)$, $(u,v)$ and $(y,y)$ for which the metric is non-zero. The latter components $(x,x)$, $(u,v)$ and $(y,y)$ yield the constraint,
\begin{equation}\label{revision}
\frac{1}{2}g_{\mu \nu}2 (1-c)\Lambda=0\, ,
\end{equation}
which means that $c=1$. Effectively, the $(u,u)$ components of the Einstein equations are,
\begin{equation}\label{expofrdiff1}
\left (\frac{\partial^2}{\partial x^2}+\frac{\partial^2}{\partial y^2}\right )G_{F}(x,y)=\frac{16\pi G p}{\frac{2\Lambda}{R_0}-1}\,\delta(x,y)T_{\mu \nu}\, ,
\end{equation}
where we used the fact that the profile function is of the form $H(u,x,y)=G_F(x,y)\delta (u)$. Hence, the profile solution is similar to the Einstein-Hilbert case, that is,
\begin{equation}\label{exposolution1}
G_{F}(x,y)=\frac{8\,G\,p}{1-\frac{2\Lambda}{R_0}}\,\ln \left( \frac{r^2}{r_0^2}\right)\, ,
\end{equation}
where $r$ is the radial coordinate in the $(x,y)$ plane, that is $r=\sqrt{x^2+y^2}$. By doing a similar analysis, the gravitational shock wave solution corresponding to the $F(R)$ gravity (\ref{oikoexp}) is,
\begin{equation}\label{oikosol}
G_{F}(x,y)= \frac{8\,G\,p}{1-\frac{B}{(A+B)^2 D}}\,\ln \left( \frac{r^2}{r_0^2}\right)\, .
\end{equation}
Notice that due to the $(x,x)$, $(u,v)$ and $(y,y)$ components of the Einstein equations, the $F(R)$ gravity is forced to satisfy the constraint $F(0)=0$, which means that in this case $C=1$.

Clearly this is because of the constraint $F(0)=0$ coming from the $(x,x)$, $(u,v)$ and $(y,y)$ components of the Einstein equations.  Let us study some other viable models, starting with the Hu-Sawicki model \cite{frhu}, in which case the $F(R)$ gravity is,
\begin{equation}\label{hufr}
F(R)=R-\mu \lambda \frac{(R/\lambda)^{2n}}{(R/\lambda)^{2n}+1},\,\,\,n,\mu,\lambda>0\, ,
\end{equation}
in which case the gravitational shock wave solution reads,
\begin{equation}\label{husolution}
G_{F}(x,y)=-4\,G\,p\,\ln \left( \frac{r^2}{r_0^2}\right)\,,
\end{equation}
which is identical to the Einstein-Hilbert solution of Eq. (\ref{einsteinsolution}). Accordingly, the Appleby-Battye model \cite{frbattye}, in which case the $F(R)$ gravity function is,
\begin{equation}\label{applebyfr}
F(R)=R-\mu \lambda \tanh (R/\lambda),\,\,\,\mu,\lambda>0,
\end{equation}
the gravitational shock wave solution is,
\begin{equation}\label{battyesolution}
G_{F}(x,y)=-\frac{4\,G\,p}{1-\mu}\,\ln \left( \frac{r^2}{r_0^2}\right)\,,
\end{equation}
which is a rescaled version of the Einstein-Hilbert solution (\ref{einsteinsolution}). Finally, the Starobinsky model \cite{frstarob}, with the $F(R)$ gravity being of the form,
\begin{equation}\label{starobfr}
F(R)=R-\mu \lambda \left [1-\frac{1}{(1+\frac{R^2}{\lambda^2})} \right],\,\,\, n,\mu, \lambda>0\, ,
\end{equation}
has the following gravitational shock wave solution,
\begin{equation}\label{starobsol}
G_{F}(x,y)=-4\,G\,p\,\ln \left( \frac{r^2}{r_0^2}\right)\, ,
\end{equation}
which is identical to the Einstein-Hilbert solution (\ref{einsteinsolution}). In Table \ref{table1} we gathered our results for the $F(R)$ gravities we mentioned above. 

As it can be seen from the solutions we obtained, the gravitational shock wave solutions are rescaled versions of the Einstein-Hilbert solution (\ref{einsteinsolution}). This result is not accidental as we now discuss.
\begin{table*}
\small
\caption{\label{table1} The gravitational shock wave solutions $H(u,x,y)=G_F(x,y)\delta (u)$ and the constraints for various cosmologically viable $F(R)$ gravities.}
\begin{tabular}{@{}crrrrrrrrrrr@{}}
\tableline
\tableline
\tableline
Cosmological Viable F(R) Gravity Model& The Gravitational Shock Wave Profile  $G_F(x,y)$
\\\tableline
$F(R)=R-2\Lambda (c-e^{-R/R_0})$ & Constraint $c=1$, and profile $G_{F}(x,y)=\frac{8\,G\,p}{1-\frac{2\Lambda}{R_0}}\,\ln \left( \frac{r^2}{r_0^2}\right)$
\\\tableline
$F(R)=R-\frac{1}{A+Be^{-R/D}}+\frac{C}{A+B}$ & Constraint $C=1$, and profile $G_{F}(x,y)= \frac{8\,G\,p}{1-\frac{B}{(A+B)^2 D}}\,\ln \left( \frac{r^2}{r_0^2}\right)$
\\\tableline
$$ &
\\\tableline
$F(R)=R-\mu \lambda \frac{(R\lambda)^{2n}}{(R\lambda)^{2n}+1}$ & $G_{F}(x,y)=-4\,G\,p\,\ln \left( \frac{r^2}{r_0^2}\right)$
\\\tableline
$F(R)=R-\mu \lambda \tanh (R/\lambda)$ & $G_{F}(x,y)=-\frac{4\,G\,p}{1-\mu}\,\ln \left( \frac{r^2}{r_0^2}\right)$
\\\tableline
$F(R)=R-\mu \lambda \left [1-\frac{1}{(1+\frac{R^2}{\lambda^2})} \right]$ & $G_{F}(x,y)=-4\,G\,p\,\ln \left( \frac{r^2}{r_0^2}\right)$
\\\tableline
\tableline
 \end{tabular}
\end{table*}

All the viable cosmological models we discussed have a similarity, since the condition $F(0)=0$ is either imposed or holds true from the beginning. So in all the above cases the following conditions hold true, $F(0)=0$ and $F'(0)\neq 0$.

 In principle the two classes of models do not by any means cover all the possible $F(R)$ gravities, but we consider here only viable models. For example, for the $R+\alpha R^2$ model \cite{starobinsky}, the gravitational shock wave solution is a simple extension of the Einstein-Hilbert solution, so we did not addressed these types of $F(R)$ gravity. A vital feature that plays a crucial role in the classification is the fact that the Ricci scalar is zero for the metric (\ref{specificmetric}), and this issue plays some role for a particular form of $F(R)$ gravities, as we evince shortly. It is worth providing the general form of gravitational shock waves solutions, for the first class of $F(R)$ gravities. Suppose that $F'(0)= \mathcal{C}_1$, then the equations of motion for the profile function $G_F(x,y)$ become,
\begin{equation}\label{profilegeneral}
\left (\frac{\partial^2}{\partial x^2}+\frac{\partial^2}{\partial y^2}\right )G_{F}(x,y)=-\frac{16\pi G p}{\mathcal{C}_1}\,\delta(x,y)\, ,
\end{equation}  
so the gravitational shock wave solution is,
\begin{equation}\label{gravitationalgeneral}
G_F(x,y)=\frac{8\pi G p}{\mathcal{C}_1}\,\ln \left( \frac{r^2}{r_0^2}\right)\, .
\end{equation}
Actually the solution (\ref{gravitationalgeneral}) is the most general non-trivial solution we can have in the case that the higher order gravity theory is an $F(R)$ gravity, except for the cases that either the $F(R)$ gravity or the first derivative $F'(R)$ are singular at $R=0$. For example, the following $F(R)$ gravities are problematic when someone seeks for gravitational shock wave solutions,
\begin{equation}\label{frsolutionsnontrivial}
F(R)=R-\alpha R^{-n},\,\,\,F(R)=R+\alpha \ln R\, ,
\end{equation}
with $n>0$. These situations obviously are different from the ones we discussed here, so this analysis is deferred to a future work, were singular situations like these will be studied.

Finally, let us discuss another non-trivial issue. Consider the $F(R)$ gravity,
\begin{equation}\label{frextracase}
F(R)=b\, e^{\Lambda R}\, ,
\end{equation}
and owing to the condition $F(0)=0$, this gravity is forced to obey $b=0$, so it yields a trivial result. In order for these gravities to yield a non-trivial result, a cosmological constant $\Lambda_1$ has to be added in their functional form, so eventually the $F(R)$ gravity would be,
\begin{equation}\label{frextracase}
F(R)=b\, e^{\Lambda R}-\Lambda_1\, ,
\end{equation}
so the condition $F(0)=0$ imposes the constraint $b=\Lambda_1$. The gravitational shock wave profile is also a rescaled version of the Einstein-Hilbert solution.

\subsection{Solutions from Various Higher Order Gravities}

Now we proceed to other higher order gravities, in which case the higher curvature invariants $\Psi$ and $\Omega$ are included in the Lagrangian. Since for the specific metric (\ref{specificmetric}), we have $\Psi=\Omega=0$, in effect all the $F(R,\Psi,\Omega)$ gravities that contain terms $R^n$, $\Psi^m$, $\Omega^k$, with $n,m,k\geq 2$, will have the gravitational shock wave solution corresponding to general relativity, given in Eq. (\ref{einsteinsolution}). In Table \ref{table2} we list some interesting cases that all have the Einstein-Hilbert gravitational shock wave solution. So we focus on the non-trivial cases, and we mainly discuss $F(R,\Psi)$ gravities for simplicity, since similar results can be obtained for the $F(R,\Omega)$ case. The non-trivial case $F(R,\Psi,\Omega)=R+\alpha R^2+b\Psi+d\Omega$, was studied in detail in Ref. \cite{gsw14}, so we do not discuss this case here. 
\begin{table*}
\small
\caption{\label{table2} Higher order $F(R,\Psi,\Omega)$ gravities for which the gravitational shock wave solution is the Einstein-Hilbert one $H(u,x,y)=-4\,G\,p\,\ln \left( \frac{r^2}{r_0^2}\right)\delta (u)$.}
\begin{tabular}{@{}crrrrrrrrrrr@{}}
\tableline
\tableline
\tableline
$F(R,\Psi,\Omega)$ Gravity
\\\tableline
$F(R,\Omega)=R+\gamma \Omega^{n},\,\,\,\,n\geq 2$
\\\tableline
$F(R,\Psi,\Omega)=R+\gamma \Omega^{n}\Psi^m,\,\,\,\,n,m\geq 2$
\\\tableline
$F(R,\Psi,\Omega)=R+\gamma \Omega^{n}+\delta \Psi^m,\,\,\,\,n,m\geq 2$
\\\tableline
$F(R,\Psi,\Omega)=f(R)+\gamma \Omega^{n}+\delta \Psi^m,\,\,\,\,n,m\geq 2,\,\,\,f(0)=0,\, f'(0)=1$
\\\tableline
$F(R,\Psi,\Omega)=f(R)+K(\Omega,\Psi),\,\,f(0)=0,\, f'(0)=1,\,\,\frac{\partial K(\Psi,\Omega)}{\partial \Psi}(0,0)=0,\,\,\,K(\Psi,\Omega)(0,0)=0$
\\\tableline
\tableline
 \end{tabular}
\end{table*}
The non-triviality in the gravitational shock wave solutions for $F(R,\Psi)$ gravities can occur if one of the following conditions hold true,
\begin{equation}\label{conditionsfrpsi}
F_R(0,\Psi)\neq 0,\,\,\,F_{\Psi}(R,0)\neq 0.
\end{equation} 
There are various forms of $F(R,\Psi)$ gravities which can satisfy all, or at least some of these conditions, for example if one chooses a gravity of the form $F(R,\Psi)=ae^{b\,R}+ce^{d\,\Psi}$, or $F(R,\Psi)=F(R)\times f(\Psi)$, and the function $F(R)$ is chosen to be one of the viable cosmological models we studied in the previous section. 

Let us explicitly demonstrate what are the solutions for a simple example satisfying one of the conditions of Eq. (\ref{conditionsfrpsi}), so we study the case $F(\Psi)=\beta e^{\Lambda \Psi}-\Lambda_1$, the cosmological implications of which, were studied in Ref. \cite{highord3}. For a general $F(\Psi)$ function and metric, the gravitational equations read,
\begin{align}\label{fpsicase}
& \frac{1}{2}g_{\mu \nu}F(\Psi)-2(F_{\Psi}R^{\alpha}_{\mu}R_{\alpha \nu}-g_{\mu \nu}\nabla_a\nabla_b(F_{\Psi} R^{ab})+2\nabla_a\nabla_{\nu}(F_{\Psi} R^{a}_{\mu})+2\nabla_a\nabla_{\mu}(F_{\Psi} R^{a}_{\nu})+8\pi G T_{\mu \nu}=0\, ,
\end{align}
so for the specific metric (\ref{specificmetric}) and for the function $F(\Psi)=\beta e^{\Lambda \Psi}-\Lambda_1$, the gravitational equations become,
\begin{equation}\label{nontrivialcase1}
 \nabla_{x,y}^4 G_{F}(x,y)=-\frac{16\pi G p}{\Lambda \beta}\delta (x,y)\, ,
\end{equation}
with $\nabla_{x,y}=\frac{\partial}{\partial x}\hat{x}+\frac{\partial}{\partial y}\hat{y}$. Notice that the $(x,x)$, $(u,v)$ and $(y,y)$ components of the field equations impose the constraint $F(0)=0$, so this means that $\beta=\Lambda_1$. The differential equation (\ref{nontrivialcase1}) is known as the biharmonic equation \cite{biharmonic}, so by using the integral,
\begin{equation}\label{integral}
\int \frac{e^{ikr}}{k^4}=\frac{\pi}{2}r^2\left(\ln r-1\right)\, ,
\end{equation}
and seeking for radially symmetric solutions, the methods of the fourier transformed Green's function yields the following solution,
\begin{equation}\label{solutionnontrivial}
G_{F}(x,y)=-\frac{G p}{\pi \Lambda \beta}r^2\left(\ln r-1\right)\, .
\end{equation}
The singularity structure of the above gravitational shock wave solution will be studied at a later section.

Another non-trivial gravitational shock wave solution results in the case that the $F(R,\Psi)$ function is of the form $F(R,\Psi)= R+\beta e^{\Lambda \Psi}-\Lambda_1$, or more generally in the case that $F(R,\Psi)=F_1(R)+F_2(\Psi)-\Lambda_1$, with $F_1(0)=0$, $F_1'(0)=1$, $F_{2}(0)\neq 0$ and $F_2'(0)\neq 0$, in which case, the gravitational equations become,
\begin{align}\label{fpsicase}
& G_{\mu \nu}=\frac{1}{2}g_{\mu \nu}(F(\Psi)-\Lambda_1)-2(F_{\Psi}R^{\alpha}_{\mu}R_{\alpha \nu}-g_{\mu \nu}\nabla_a\nabla_b(F_{\Psi} R^{ab})+2\nabla_a\nabla_{\nu}(F_{\Psi} R^{a}_{\mu})+2\nabla_a\nabla_{\mu}(F_{\Psi} R^{a}_{\nu})+8\pi G T_{\mu \nu}\, ,
\end{align}
so for the metric (\ref{specificmetric}) and for $R+\beta e^{\Lambda \Psi}-\Lambda_1$, the gravitational equations become,
\begin{equation}\label{highlynontrivial}
\left ( \Lambda \beta \nabla_{x,y}^4+\nabla_{x,y}^2\right) G_F(x,y)=-16 \pi G p \delta (x,y)\, .
\end{equation}
Notice that in this case, the $(x,x)$, $(u,v)$ and $(y,y)$ components of the gravitational equations yield the constraint $\beta =\Lambda_1$.  This differential equation has been solved in Ref. \cite{gsw15}, so the solution is,
\begin{equation}\label{newsolutionrev}
G_F(x,y)=-8\,G\,p\left( K_0(\frac{r}{\sqrt{-\beta \Lambda}})+\ln (\frac{r}{r_0})\right)\, .
\end{equation}

In general, this class of solutions is obtained from $F(R,\Psi)$ gravities of the form,
\begin{equation}\label{peculiarcomplexcase}
F(R,\Psi)=F_1(R)+F_2(\Psi),\,\,\,F_1(0)=C_1,\,F_1'(0)=C_2,\,F_2(0)=C_3,\,F_2'(0)=C_4\, ,
\end{equation}
with $C_i$, $i=1,...4$, being constants. The resulting gravitational equations in this case are,
\begin{equation}\label{highlynontrivial}
\left ( C_3C_4 \nabla_{x,y}^4+C_2\nabla_{x,y}^2\right) G_F(x,y)=-16 \pi G p \delta (x,y)\, ,
\end{equation}
since the $(x,x)$, $(u,v)$ and $(y,y)$ components of the gravitational equations impose the constraint $C_1=-C_3$. A similar gravitational shock wave solution to the general relativistic one, can be obtained in the case that $F(R,\Psi)=R\gamma e^{\Lambda \Psi}$, in which case the gravitational shock wave solution is a rescaled version of solution (\ref{einsteinsolution}), which is,
\begin{equation}\label{avak}
G_F(x,y)=-\frac{8Gp}{\gamma} \ln \frac{r}{r_0}\, .
\end{equation}
Finally, we need to note that it is conceivable that any combination of the $F(R)$ gravities we studied in the previous section, with polynomials of $\Psi$ and $\Omega$, that is,
\begin{align}\label{finalformsfunctions}
& F(R,\Psi,\Omega)=f(R)+\gamma \Omega ^n+\beta \Psi^m,\\ \notag &
F(R,\Psi,\Omega)=f(R)+\gamma \Omega ^n\beta \Psi^m\, ,
\end{align}
with $n,m\geq 2$ will yield the solutions of the simple $f(R)$ gravity case which we studied in the previous section. So practically the higher polynomials of the Kretschmann scalar and the Ricci tensor squared, have no effect on the gravitational shock wave solutions. 

\subsubsection{The Gauss-Bonnet Gravity Case}

Having discussed the solutions of gravitational shock waves in the context of the higher order gravities, in this section we study the case of an $R+F(\mathcal{G})$ higher order gravity, with $\mathcal{G}=R^2-4R_{\mu \nu}R^{\mu \nu}+R_{\mu \nu \lambda k}R^{\mu \nu \lambda k}$ being the Gauss-Bonnet scalar. Gauss-Bonnet modified gravity theories have been studied both in cosmological and astrophysical contexts, see for example \cite{gaussb1,gaussb2,gaussb3,gaussb4,lobo,myrzafgfinite}. In this section we are interested to find the gravitational shock wave solutions for several realistic examples of $R+F(\mathcal{G})$ gravities. Particularly, we shall discuss in some detail the following models,
\begin{equation}\label{cand1}
F(\mathcal{G})=\frac{a_1\mathcal{G}^n+b_1}{a_2\mathcal{G}^n+b_2}\, ,
\end{equation}
\begin{equation}\label{cand2}
F(\mathcal{G})=\frac{a_1\mathcal{G}^{n+N}+b_1}{a_2\mathcal{G}^n+b_2}\, ,
\end{equation}
\begin{equation}\label{cand3}
F(\mathcal{G})=a_3\mathcal{G}^n (b_3\mathcal{G}^m+1)\, ,
\end{equation}
\begin{equation}\label{cand4}
F(\mathcal{G})=\mathcal{G}^m\frac{a_1\mathcal{G}^n+b_1}{a_2\mathcal{G}^n+b_2}\, ,
\end{equation}
with the parameters $a_i$, and $b_i$, $i=1,2,3$, being arbitrary real constants, and $n>1$, $N>0$, $m>0$. Note that these types of $F(\mathcal{G})$ are known to exhibit cosmological finite time singularities \cite{Nojiri:2005sx,oiksing}, but here we shall reveal another astrophysical aspect of these modified gravity models.

The gravitational action of a general $R+F(\mathcal{G})$ theory is \cite{reviews1,reviews1a,gaussb1,gaussb2,gaussb3,gaussb4,lobo,myrzafgfinite},
\begin{equation}\label{actionfggeneral}
\mathcal{S}=\frac{1}{2\kappa^2}\int \mathrm{d}x^4\sqrt{-g}\left [ R+F(\mathcal{G})\right ]+S_m,
\end{equation}
with $\kappa^2=8\pi G$ denoting the gravitational constant and $S_m$ stands for the matter content of the theory at hand, with energy momentum tensor $T_{\mu \nu}$. Upon variation of the action with respect to the metric tensor, the gravitational equations of motion easily follow,
\begin{eqnarray}
\label{fgr1}
&& \!\!\!\!\!\!\!\!\!\!
G_{\mu \nu}-\frac{1}{2}g_{\mu \nu}F(\mathcal{G})+\left(2RR_{\mu \nu}-4R_{\mu 
\rho}R_{\nu}^{\rho}+2R_{\mu}
^{\rho \sigma \tau}R_{\nu \rho \sigma \tau}-4g^{\alpha \rho}g^{\beta \sigma}R_{\mu \alpha 
\nu \beta}
R_{\rho \sigma}\right)F'(\mathcal{G})\notag 
\\ 
&& \  +4 \left[\nabla_{\rho}\nabla_{\nu}F'(\mathcal{G})\right ] R_{\mu}^{\rho}
-4g_{\mu \nu} \left [\nabla_{\rho}\nabla_{\sigma }F'(\mathcal{G})\right ]R^{\rho \sigma }+4 \left 
[\nabla_{\rho}\nabla_{\sigma }F'(\mathcal{G})\right ]g^{\alpha \rho}g^{\beta \sigma }R_{\mu \alpha 
\nu \beta }
\notag 
\\ 
&& \
-2 \left [\nabla_{\mu}\nabla_{\nu}F'(\mathcal{G})\right ]R+2g_{\mu \nu}\left [\square F'(\mathcal{G}) 
\right]R
\notag 
\\
&&\
-4 \left[\square F'(\mathcal{G}) \right ]R_{\mu \nu }+4 
\left[\nabla_{\mu}\nabla_{\nu}F'(\mathcal{G})\right]R_{\nu}^{\rho }
=\kappa^2T_{\mu \nu }\, .
\end{eqnarray}
A crucial feature in our analysis, that will eventually determine the final form of the gravitational shock wave solution, is the fact that the Gauss-Bonnet scalar is zero for the metric (\ref{specificmetric}). This simplifies to a great extent the final picture as we now demonstrate, and practically the gravitational shock wave solutions for the $F(\mathcal{G})$ models (\ref{cand1}), (\ref{cand2}), (\ref{cand3}) and (\ref{cand4}) have similar gravitational shock wave solutions. 

Let us see this explicitly, so for the models (\ref{cand1}) and (\ref{cand2}), we have $F(0)= \frac{b_1}{b_2}$ and $F'(0)=0$, and therefore the gravitational equations of motion (\ref{fgr1}) become,
\begin{equation}\label{profilegeneralgaussbonnet}
\left (\frac{\partial^2}{\partial x^2}+\frac{\partial^2}{\partial y^2}\right )G_{F}(x,y)=-16\pi G p\,\delta(x,y)\, .
\end{equation}
Notice that for deriving the resulting gravitational equations (\ref{profilegeneralgaussbonnet}), we also took into account the $(x,x)$, $(u,v)$ and $(y,y)$ components of the gravitational equations, which result in the constraint $F(0)=0$ which for the models (\ref{cand1}) and (\ref{cand2}) implies that $b_1=0$. Also in this case too we assumed that $T_{uu}=p\,\delta(x,y) \delta (u)$, so the shock wave is generated by a massless ultra-relativistic particle mass source, with momentum $p$. The gravitational shock wave solution in this case is,
\begin{equation}\label{gravitationalgeneralgaussbonnet}
G_F(x,y)=-4\,G\,p\,\ln \left( \frac{r^2}{r_0^2}\right)\, ,
\end{equation}
with $r=\sqrt{x^2+y^2}$, and which is identical to the Einstein-Hilbert solution. 
\begin{table*}
\small
\caption{\label{table3} The gravitational shock wave solutions $H(u,x,y)=G_F(x,y)\delta (u)$, for realistic $F(\mathcal{G})$ gravities, and the imposed constraints in their functional form.}
\begin{tabular}{@{}crrrrrrrrrrr@{}}
\tableline
\tableline
\tableline
Form of the $R+F(\mathcal{G})$ Gravity Model& The Gravitational Shock Wave Profile  $G_F(x,y)$
\\\tableline
$F(R,\mathcal{G})=R+\frac{a_1\mathcal{G}^n+b_1}{a_2\mathcal{G}^n+b_2}$ & $b_1=0$ and solution $G_{F}(x,y)=-4\,G\,p\,\ln \left( \frac{r^2}{r_0^2}\right)$
\\\tableline
$F(R,\mathcal{G})=R+\frac{a_1\mathcal{G}^{n+N}+b_1}{a_2\mathcal{G}^n+b_2}$ & $b_1=0$ and solution $G_{F}(x,y)= -4\,G\,p\,\ln \left( \frac{r^2}{r_0^2}\right)$
\\\tableline
$F(R,\mathcal{G})=R+a_3\mathcal{G}^n (b_3\mathcal{G}^m+1)$ & $G_{F}(x,y)=-4\,G\,p\,\ln \left( \frac{r^2}{r_0^2}\right)$
\\\tableline
$F(R,\mathcal{G})=R+\mathcal{G}^m\frac{a_1\mathcal{G}^n+b_1}{a_2\mathcal{G}^n+b_2}$ & $G_{F}(x,y)=-4\,G\,p\,\ln \left( \frac{r^2}{r_0^2}\right)$
\\\tableline
\tableline
 \end{tabular}
\end{table*}
The other two $F(\mathcal{G})$ models, namely (\ref{cand3}) and (\ref{cand4}), have the property that $F(0)=0$ and $F'(0)$, so the gravitational equations in this case become,
\begin{equation}\label{einteincaseequationgaussbonnet}
\left (\frac{\partial^2}{\partial x^2}+\frac{\partial^2}{\partial y^2}\right )G_{EH}(x,y)=-16\pi G p\,\delta(x,y)\, ,
\end{equation}
with the solution to this differential equation being identical to the Einstein-Hilbert gravitational shock wave solution of Eq. (\ref{einsteinsolution}). In conclusion, even in the case that realistic Gauss-Bonnet modified gravities are considered, the gravitational shock waves solutions are identical to the Einstein-Hilbert solution. The results for the Gauss-Bonnet models are gathered in Table \ref{table3}. It is conceivable that the list is not exhaustive, since there exist models that may contain inverse powers of the Gauss-Bonnet scalar, or even terms proportional to $\ln \mathcal{G}$, which could potentially generate problems, however the focus in this section was on well-known and realistic $F(\mathcal{G})$ theories.

\subsection{Brief Comparison of the Gravitational Shock Wave Profile Singularity Structure}

\begin{figure}[h] \centering
\includegraphics[width=15pc]{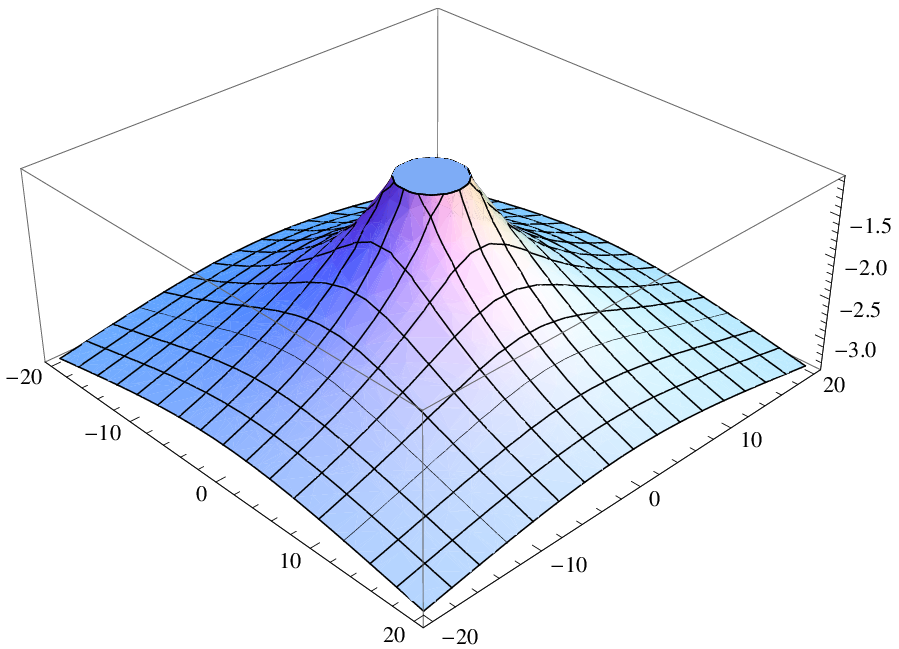}
\includegraphics[width=15pc]{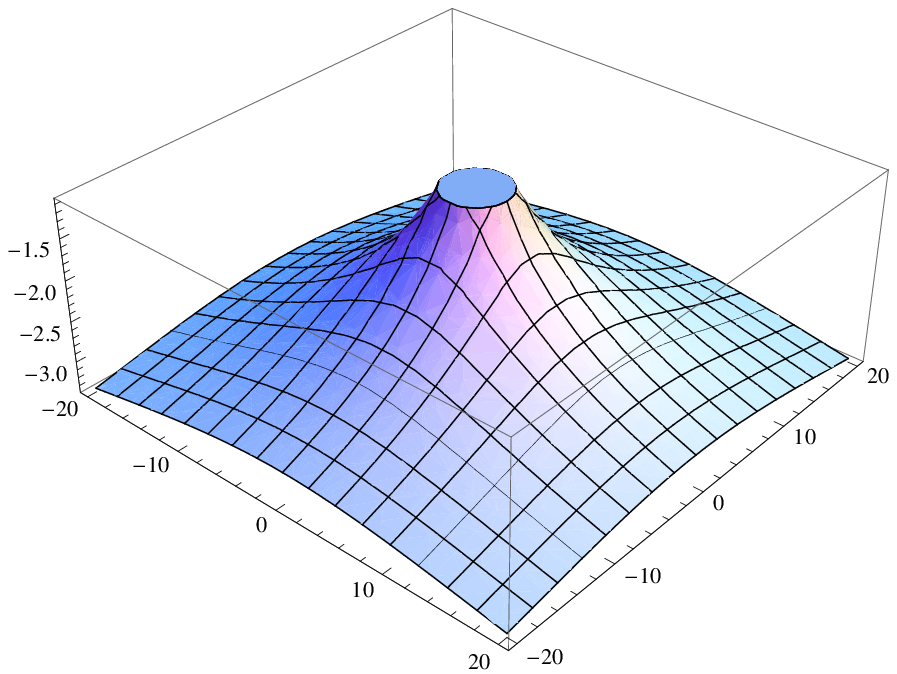}
\includegraphics[width=15pc]{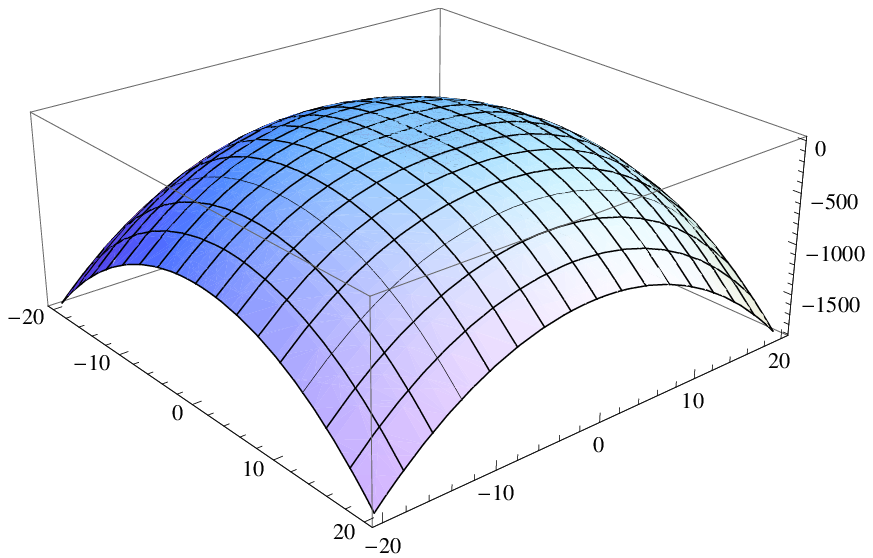}
\caption{The gravitational shock wave profiles for higher order gravities. The left plot corresponds to profile I, the right plot to profile II and the bottom plot to profile III, listed in Table \ref{table4}.}
\label{plot1}
\end{figure}

The main result we obtained in the previous sections is that the profile function of the gravitational shock wave in the context of realistic higher order gravities, has three different forms, which we list in Table \ref{table4}. As it can be seen, the profile I is the general relativistic profile, the profile II appears only in specific forms of $F(\Psi,R)$ gravities, and the profile III occurs only for $F(R_{\mu \nu}R^{\mu \nu})$ theories, and only in the case that $F_{\Psi}(0)\neq 0$. 
\begin{table*}[h]
\small
\caption{\label{table4} Three Categories of the Most Frequently Occurring Gravitational Shock Wave Profile Functions $H(u,x,y)=G_F(x,y)\delta (u)$, for Realistic Higher Order Gravities.}
\begin{tabular}{@{}crrrrrrrrrrr@{}}
\tableline
\tableline
Profile I & $G_{F}(x,y)=-4\,G\,p\,\ln \left( \frac{r^2}{r_0^2}\right)$
\\\tableline
Profile II & $G_F(x,y)=-8\,G\,p\left( K_0(\frac{r}{\sqrt{-\beta \Lambda}})+\ln (\frac{r}{r_0})\right)$
\\\tableline
Profile III & $G_{F}(x,y)=-\frac{G p}{\pi \Lambda \beta}r^2\left(\ln r-1\right)$
\\\tableline 
\tableline
 \end{tabular}
\end{table*}
The most peculiar case corresponds to the profile III, and the plot of the profile can be seen in the bottom plot of Fig. \ref{plot1}. In this case, the $r=0$ singularity seems not to occur, in contrast to the profile I but also with profile II cases.
The profiles I and II have similar properties, as it can be seen by left and right plots of Fig. \ref{plot1}, but it is worth having a more quantitative picture, so we expand the profile function I in the limit $r\to 0$, and we get,
\begin{align}\label{asymptoticlimit}
& -K_0(r)-\ln r\simeq (\gamma-ln 2)+\frac{1}{4} (-1+\gamma-\ln 2+\ln r) r^2,\\ \notag &
\, ,
\end{align}
where $\gamma$ is the Euler gamma number. The rest two profiles have identical expansion with their functional form. By taking the limits $r\to 0$, it can be seen that the singularities of the profiles II and III are milder, and in fact the profile III yields $\lim _{r\to 0}r^2(\ln r-1)=0$. 

In conclusion the effects of higher order gravity to the singularity structure of the gravitational shock wave profile are important, since the profile function approaches the $r=0$ singularity in a less steeper way for some higher order gravity theories, and this feature has also been pointed out in \cite{gsw14}. More importantly, for higher order gravity theories of the form $F(R_{\mu \nu}R^{\mu \nu})$ which satisfy $F_{\Psi}(0)\neq 0$, no singularity occurs at $r=0$, and we believe this is a significant difference between the Einstein-Hilbert solution and the higher order gravity solutions.

\section{Conclusions}

In this work we studied gravitational shock waves solutions in the context of higher order gravities. Particularly, we focused on higher order gravities of the form $F(R,R_{\mu \nu}R^{\mu \nu}, R_{\mu \nu k \lambda }R^{\mu \nu k \lambda})$ and also to Gauss-Bonnet theories of gravity of the form $R+F(\mathcal{G})$. In the case of $F(R)$ gravity, we investigated which are the gravitational shock waves solutions corresponding to various cosmologically viable theories and we found that the gravitational shock waves solutions are similar to the Einstein-Hilbert solutions. The same picture occurs also in the case of $R+F(\mathcal{G})$ gravity, when realistic gravities are taken into account. The same solutions also appear in the case of various combinations of $F(R,R_{\mu \nu}R^{\mu \nu}, R_{\mu \nu k \lambda }R^{\mu \nu k \lambda})$ gravities. Notably, polynomial functional forms of the Ricci squared tensor or the Kretschmann scalar give no contribution to the gravitational shock wave solution. A highly non-trivial gravitational shock wave solution results for an $F(R_{\mu \nu}R^{\mu \nu})$ gravity, which satisfies $F_{\Psi}(0)\neq 0$, with $\Psi=R_{\mu \nu}R^{\mu \nu}$. This particular solution has the appealing property that it has no singularity at $r=0$, in contrast to the other two solutions we found.

The study we performed is not exhaustive, meaning that there are more possibilities of choosing the higher order gravities. However, our study was devoted in well known viable and realistic gravities, and we found that there are three classes of solutions. An interesting task would be to study the gravitational shock wave solutions in the case of $F(R,T)$ gravities \cite{saridakis,capp1}, or even more complicated forms of $F(R,\mathcal{G})$ gravities \cite{cappofrg}. In principle, other non trivial solutions might appear too. 

Another interesting issue that could formally be addressed in a future work is to find a rigid interpretation for the cases that the higher order gravities functional forms are singular at $R=0$ or equivalently at $\Psi=\Omega=0$, with $\Omega$ being the Kretschmann scalar. We discussed in brief some cases in the previous sections, and this study should be done in detail in a future work.

Also, the study we performed assumed that the ultra-relativistic particle that generates the gravitational shock wave background, propagates in a Minkowski background, so it would be interesting to examine the higher order gravitational shock wave solutions for the case that the particle propagates in a curved background, like one of the black holes backgrounds. In the latter case, it would be interesting to see if the Hawking radiation effect is affected from these ultra-relativistic propagating particles, always in the context of higher order gravities.

Finally, and in relation to propagation in curved backgrounds, it is worth studying the effects of higher order gravities in collisions of gravitational shock waves. Some studies in the past were devoted in this issue, see for example \cite{gsw9,gsw10}, so it would be interesting to extend these works in the context of higher order gravity. Also the cosmological implications of gravitational shock waves are also studied in the literature \cite{gsw11}, so an interesting study would be to seek for cosmological implications of gravitational shock waves in the case that the waves originate from a higher order gravity.

\section*{Acknowledgments}

This work is supported by Min. of Education and Science of Russia (V.K.O).

\section*{Appendix: Christoffel Symbols and Curvature Tensors of the Gravitational Shock Wave Metric}

Here we present in detail the Christoffel symbols and the components of the Riemann and Ricci tensors corresponding to the gravitational shock wave metric, 
\begin{equation}\label{specificmetricappendix}
\mathrm{d}s^2=-\mathrm{d}u\mathrm{d}v+H(u,x,y)\mathrm{d}u^2+\mathrm{d}x^2+\mathrm{d}y^2\, .
\end{equation}
The Christoffel symbols are given below,
\begin{align}\label{christf}
& \Gamma^2_{1\,1}=-\partial_u\,H(u,x,y),\,\,\,\Gamma^2_{3\,1}=-\partial_x\,H(u,x,y),\,\,\,\Gamma^2_{4\,1}=-\partial_y\,H(u,x,y),\\ \notag &\,\,\,\Gamma^3_{1\,1}=-\frac{1}{2}\partial_x\,H(u,x,y),\,\,\,\Gamma^4_{1\,1}=-\frac{1}{2}\partial_y\,H(u,x,y)\, .
\end{align}
The only non-zero component of the Ricci tensor $R_{\mu \nu}$ is,
\begin{equation}\label{Ruu}
R_{uu}=-\frac{1}{2}\left (\frac{\partial^2}{\partial x^2}+\frac{\partial^2}{\partial y^2}\right )H(u,x,y)\, .
\end{equation}
The Ricci scalar $R$, the Ricci tensor squared $R_{\mu \nu }R^{\mu \nu}$, the Riemann tensor squared $R_{\mu \nu k \lambda}R^{\mu \nu k \lambda}$ and the Gauss-Bonnet scalar, calculated for the metric (\ref{specificmetricappendix}), are equal to zero, that is,
\begin{equation}\label{highercurvatures}
R=0,\,\,\,R_{\mu \nu }R^{\mu \nu}=0,\,\,\,R_{\mu \nu k \lambda}R^{\mu \nu k \lambda}=0,\,\,\,\mathcal{G}=R^2-4R_{\mu \nu}R^{\mu \nu}+R_{\mu \nu \lambda k}R^{\mu \nu \lambda k}=0\, .
\end{equation}
Finally, the non-zero components of the Riemann tensor are the following,
\begin{align}\label{riemanntensor}
& R^{2}_{3\,1\,3}=\partial_{x}^2H(u,x,y),\,\,\,R^{2}_{3\,1\,4}=\partial_{(x,y)}H(u,x,y),\,\,\,R^{2}_{3\,3\,1}=-\partial_{x}^2H(u,x,y),\,\,\,\\ \notag &
R^{2}_{3\,4\,1}=-\partial_{(x,y)}H(u,x,y),\,\,\,R^{2}_{4\,1\,3}=\partial_{y}H(u,x,y),\,\,\,R^{2}_{4\,1\,4}=\partial_{y}^2H(u,x,y),\,\,\,\\ \notag &
R^{2}_{4\,3\,1}=-\partial_{(x,y)}H(u,x,y),\,\,\,R^{2}_{4\,4\,1}=-\partial_{y}^2H(u,x,y),\,\,\,R^{3}_{1\,1\,3}=\frac{1}{2}\partial_{x}^2H(u,x,y),\,\,\,\\ \notag &
R^{3}_{1\,1\,4}=\frac{1}{2}\partial_{(x,y)}H(u,x,y),\,\,\,R^{3}_{1\,3\,1}=-\frac{1}{2}\partial_{x}^2H(u,x,y),\,\,\,R^{3}_{1\,4\,1}=-\frac{1}{2}\partial_{(x,y)}H(u,x,y),\,\,\,\\ \notag &
R^{4}_{1\,1\,2}=\frac{1}{2}\partial_{(x,y)}H(u,x,y),\,\,\,R^{4}_{1\,1\,4}=\frac{1}{2}\partial_{y}^2H(u,x,y),\,\,\,R^{4}_{1\,3\,1}=-\frac{1}{2}\partial_{(x,y)}H(u,x,y),\,\,\,\\ \notag &
R^{4}_{1\,4\,1}=-\frac{1}{2}\partial_{y}^2H(u,x,y)
\end{align}

\end{document}